# The Structure of Signals: Causal Interdependence Models for Games of Incomplete Information

**Michael P. Wellman**
Computer Science & Engineering
University of Michigan
*wellman@umich.edu*

**Lu Hong**
Finance
Loyola University Chicago
*lhong@luc.edu*

**Scott E. Page**
Political Science
University of Michigan
*spage@umich.edu*

## Abstract

Traditional economic models typically treat private information, or *signals*, as generated from some underlying state. Recent work has explicated alternative models, where signals correspond to interpretations of available information. We show that the difference between these formulations can be sharply cast in terms of causal dependence structure, and employ graphical models to illustrate the distinguishing characteristics. The graphical representation supports inferences about signal patterns in the interpreted framework, and suggests how results based on the generated model can be extended to more general situations. Specific insights about bidding games in classical auction mechanisms derive from qualitative graphical models.

## 1 Introduction

In a game of incomplete information, we commonly distinguish the information available to respective agents in terms of their *epistemic types* (henceforth just *types*). Agents are assumed to know their own private types, and have probabilistic beliefs about other-agent types. When other-agent types are strategically relevant, our analyses generally consider what agents can infer about others from their own types, a possibility that arises only when types are *interdependent*. Such interdependencies typically derive from the way that the respective agents' information relates to some underlying state of the world, in which case we may refer to an agent's type as a *signal* about that state.

The treatment of signals is a central element of theoretical analysis for many classes of multiagent interactions. In particular, signals play a pivotal role in formal models of auctions, voting, and many other social mechanisms. Canonical theoretical results about these problems take the form of relating properties of the signal distributions to solutions (e.g., equilibria) of the Bayesian games induced. For example, celebrated theorems in the auction literature establish revenue comparisons (ascending open outcry greater than second-price sealed bid greater than first-price sealed bid) under the assumption that signals are *affiliated* (a form of monotone relation) [Milgrom and Weber, 1982].

Recent work by Hong and Page [2009] has shown that the implications of signals on interacting agents varies qualitatively depending on whether they are viewed as caused by payoff-relevant elements of the underlying state (*generated*) or as causal factors from a predictive model of the observer (*interpreted*). Although both represent common patterns of uncertain information, literature to date has tended to emphasize the generated view.

We develop a graphical model framework for representing the *structure* of signals in games of incomplete information. We show that this representation can capture the key structural distinctions between generated and interpreted signals, and supports graphical inference to identify the relevant patterns in complex information relationships. As a modeling language, the approach also supports representation of complex games of information, and automated game-theoretic inference about these games. Observations highlighted in the sections below show several ways in which modeling the structure of signals supports generalized insights about classes of incomplete-information games.

## 2 Representation of Information Structure

### 2.1 Bayesian Games

In a game of incomplete information, some strategically relevant aspects of the game are not known to at least some agents. To play or predict the result of such a game, it is essential to model what private information agents have about these uncertain aspects. In his seminal work, Harsanyi [1967] used *types* to capture agents' private information and defined a game where agents' payoff functions depend on the types of all agents. It is assumed that each agent knows her own type but not others. Her beliefs about other agents' types are given by a probability distribution over possible types of other agents conditional on her own type. Harsanyi called this a *Bayesian game*. The Bayesian game formulation has since become the standard framework for games of incomplete information.

The treatment of types in Bayesian games is useful and elegant, but provides no account of the *origins* of private information. That agents' types may arbitrarily interdepend serves generality, but may abstract away structure in *how* they interdepend, which in turn may support qualitatively significant distinctions. So how can we incorporate into the framework where private information comes from? We posit a set of underlying states of nature, $\Omega$, such that all relevant uncertainty can be captured in terms of which state $\omega \in \Omega$ obtains. This includes state relevant to payoff functions (which we refer to as the *outcome*, $v$), as well as state bearing on the production of types. Private information, then, can be characterized as a *signal* providing evidence about $\omega$. From a particular agent $i$'s perspective, a signal $s_i$ drawn from its set of possible signals $S_i$ yields information about the underlying state according to the joint probability of signal and state, $\Pr(s_i; \omega)$.[1] The relationship among the signals of $N$ agents is captured by the joint conditional distribution $\Pr(s; \omega)$, where $s = (s_1, \ldots, s_N)$. One interesting case is conditional independence, defined by the equality $\Pr(s_i \mid s_{-i}; \omega) = \Pr(s_i \mid \omega)$, where $s_{-i} = (s_1, \ldots, s_{i-1}, s_{i+1}, \ldots, s_N)$ denotes the signals of all agents other than $i$.

The goal of capturing structure in the origin of signals is now reduced to capturing structure in a particular joint probability distribution. For that we appeal to the formalism of graphical models.

## 2.2 Graphical Models

Probabilistic graphical models [Koller and Friedman, 2009, Pearl, 1988] express a distribution over world states, factored across a product space of random variables. The defining feature of such models is their explicit representation of probabilistic dependence in terms of graph structure.[2] Formally, a Bayesian network is an annotated directed acyclic graph, comprising:

1. nodes $\{X_1, \ldots, X_n\}$ representing $n$ random variables,
2. directed edges between nodes, with $Pa(X)$ denoting the *parents* of $X$: nodes with a directed edge to $X$, and
3. a conditional probability function $\Pr(X_i \mid Pa(X_i))$ associated with each node $i = 1, \ldots, n$.

The graph structure encodes the pattern of conditional independence relations that follow from the fundamental assumption that each random variable is conditionally independent of its non-descendants (nodes to which it has no directed path) given its parents. Under a causal interpretation of edges in the graph, the Bayes net also provides a representation suitable for inference about decisions and experimentation [Pearl, 2009]. From the independence semantics of the graphical model, we can reconstruct the full joint distribution as the product of conditional probabilities directly represented:

$$\Pr(x_1, \ldots, x_n) = \prod_{i=1}^{n} \Pr(x_i \mid Pa(X_i)).$$

The factored representation affords significant benefits in compact expression of joint probability distributions, as well as algorithmic efficiencies in inference.

Moreover, the graph itself supports reasoning about conditional independence. For instance, it follows that a node is conditionally independent of all others given its *Markov blanket*: parents, children, and parents of children. More generally, the Bayes-net graph entails that the set of variables $Y$ is conditionally independent of the set $Y'$ given variables in $Z$ if and only if (iff) $Z$ *d-separates* $Y$ from $Y'$ [Pearl, 1988]. Such d-separation holds when every undirected path between nodes $y \in Y$ and $y' \in Y'$ is *blocked* by $Z$. A path is blocked by $Z$ if it includes a node $z \in Z$ with at least one edge pointing out, or a *collider* node with both edges pointing in such that neither the collider nor any of its descendants is in $Z$.

## 2.3 Example: Interdependent Values in Auction Games

To illustrate the use of graphical models for expressing dependencies in signal structure, we present Bayes nets capturing some basic patterns of interdependent values in auction games. Some classic results in auction theory [Krishna, 2010], such as Vickrey's original analysis [Vickrey, 1961], assume *independent private values* (IPV): that the agents' types (valuations for a good under auction in this setting) are probabilistically independent. Auction theory recognizes, however, that in general, an agent's value for a good may depend on others' valuations. For example, the valuation of an expert (e.g., if the good is fine art or antique furniture) could sway the value of others, because it provides evidence about fundamental shared values in authenticity, craftsmanship, and other uncertain features of the good. This has direct implications for bidding strategy, in three ways. First, in an auction where information is revealed over time (e.g., the ascending open outcry), an agent should interpret the observed bidding behavior of others as evidence about their valuations. Second, the agent should anticipate the effect of its own bid on information revealed to others. Third, even when no information is revealed (e.g., in a sealed-bid auction), the others' bids will reflect their valuations, hence the agent should consider how the auction result conditional on those bids influences the probability and expected value of winning.

---
[1] Throughout we adopt notation for discrete probability and finite summations, and ask the reader to recognize that the fundamental concepts generalize to density functions and integrals.

[2] See the cited references for tutorial and in-depth treatments.

One standard model of interdependent values posits an underlying state variable, $\omega$, which bears on every agent's valuation (in the antiques example, $\omega$ might represent authenticity or craftsmanship). To model the fact that an agent $i$ has only partial information about its own valuation (leaving room for the influence of others), we define the agent's private information, $s_i$, as a *signal* that depends probabilistically on its valuation. Figure 1(a) depicts this situation as a Bayesian network. The dependence of $v_i$ on $v_j$, $j \neq i$, is represented by the undirected path between these variables through $\omega$. Node $\omega$ blocks this sole path, thus by the semantics of Bayesian networks, the model also entails that $v_i$ and $v_j$ are *conditionally independent* given $\omega$. By the same logic, the agents' signals are also marginally dependent but conditionally independent.

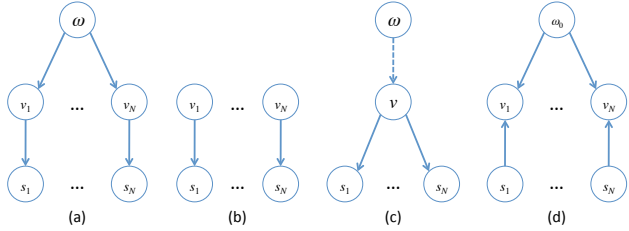

Figure 1: Interdependent values over $N$ agents, expressed in Bayesian network structures. We distinguish agent $i$'s private information (signal), $s_i$, from its valuation, $v_i$. (a) Standard model, where interdependencies are captured in a hidden state variable, $\omega$. (b) Independent private values expressed in this framework. (c) Common values as a special case of the interdependent value model. (d) Signals as cause rather than observed effect of value.

For comparison, Figure 1(b) presents the IPV model using this graphical notation. Since the expected value of $v_i$ is independent of $s_j$, $j \neq i$, the signal $s_i$ is a sufficient statistic, and assuming risk neutral preferences, we take $E[v_i \mid s_i]$ as agent $i$'s valuation. At the other extreme, we can also express the *common value* model, Figure 1(c), where every agent has the same valuation, $\forall i. v_i = v$, but observes an independent signal conditional on that value. In this case, the underlying state variable $\omega$ is superfluous, as the interdependence can be captured in terms of $v$.

The model of Figure 1(d) reverses the direction of arrows between values and signals compared to our standard model (a). In this structure, we interpret the signals as factors causing or determining the value, rather than as observed effects of these values as in the other models. For example, there may be common uncertain factors (represented by $\omega_0$) relevant to all agents' value, and each makes an observation about a feature of interest only to them. In terms of the Bayesian network, the paths between $s_i$ remain blocked by $\omega_0$, but are now *un*blocked by the $v_i$, since these are collider nodes with both arrows pointing in. Hence in model (d) the signals $s_i$ are marginally independent, but conditionally dependent given the values $v_i$, yet still rendered conditionally independent given $\omega_0$.

These differences in dependence structure have significant implications for how agents should reason about their own value given information reflecting the other agents' values—such as observations of their bids or indirect evidence about their bids from the resolution of who wins the auction. In model (a), the agent cares about other-agent signals because they influence their bids (hence the agent's own probability of winning with its own bid), and also because they are evidence for the underlying state $\omega$. In model (d), the other-agent signals provide no relevant information about the underlying state or agent's own value.

To see this, we consider a generic form of bidding game, where agent $i$'s bid $b_i(s_i)$ is a function of its signal. Its payoff (*utility*), $u_i$, is a function of its value, and of the bids of all the agents.

**Definition 1.** *A utility function $u_i(v_i, b_i, b_{-i})$ is value-action separable if it can be expressed in the form*

$$u_i(v_i, b_i, b_{-i}) = f(v_i, b_i)g(b_i, b_{-i}) + h(b_i, b_{-i}),$$

*for some functions $f$, $g$, and $h$.*

Note that payoff functions for standard auction games are decomposable in this way (as are many other games analogously), with function $g$ determining the auction winner (good allocation), and $h$ the associated payment. For example, the first-price sealed-bid auction has $f(v_i, b_i) = v_i$, $g(b_i, b_{-i}) = \frac{1}{|\{j | b_j = b_i\}|}$ if $b_i = \max_j b_j$, zero otherwise, and $h(b_i, b_{-i}) = b_i g(b_i, b_{-i})$.

**Theorem 1.** *For a game with value-action separable payoff functions, the signal model of Figure 1(d) is strategically equivalent to the IPV case (b).*

*Proof:* The objective of each agent is to choose $b_i(s_i)$ so as to maximize expected utility, given other-agent bidding strategies.

$$\mathbb{E}_{\omega_0, s_{-i} | s_i}[f(v_i, b_i)g(b_i, b_{-i}) + h(b_i, b_{-i})]$$

Since $b_i(s_i)$ is agent $i$'s choice, by d-separation of $v_i$ from $s_{-i}$ we have that $(v_i, b_i)$ and $(b_i, b_{-i})$ are conditionally independent given $s_i$. Accounting for this and other independence, the expectation can be decomposed as

$$\mathbb{E}_{\omega_0}[f(v_i, b_i)]\mathbb{E}_{s_{-i}}[g(b_i, b_{-i})] + \mathbb{E}_{s_{-i}}[h(b_i, b_{-i})].$$

Since $\omega_0$ affects only the term for agent $i$, we can marginalize it out and obtain the same problem faced by an agent under the IPV model. □

The standard auction-theoretic treatment identifies the signal $s_i$ with $v_i \mid s_i$, the agent's value conditional on signal. This is convenient since it associates the private information directly with payoff conditional on winning. It is also without loss of generality, in that the only relevance of the signal in these auction settings is its relation to value. This treatment does entail loss of structural information, however, as once we marginalize the $s_i$, the model of Figure 1(a) is indistinguishable from model (d). As argued above, this sacrifices a potential opportunity to exploit independence of signals to simplify the characterization of bidding strategy.

**Observation 2.** *Independence of value from other-agent signals may hold even given interdependent values, significantly broadening the conditions where IPV-like bidding strategies can be applied.*

## 3 The Origin of Signals

### 3.1 Generated and Interpreted Signals

Hong and Page [2009] distinguish between *interpreted* and *generated* signals. The distinction hinges on the relation of signals to a focal variable, $v$, which we refer to as the variable of interest or *outcome*. In some cases there may be individual outcome variables for each agent, $v_1, \ldots, v_N$. The letter "v" evokes the value variable in our auction example (§2.3), but this is just one special case.

Generated signals are often motivated in terms of imprecise measurements or distortions of an outcome variable $v$. The example of Figure 1(a) is an instance of generated signals. A generated signal is characterized by its distribution conditional on the variable of interest, $\Pr(s \mid v)$. When signals for respective agents are conditionally independent given outcomes (or individualized outcomes), we have for all $i$, $\Pr(s_i \mid v_i, s_{-i}) = \Pr(s_i \mid v_i)$, which is exactly the information required by the Bayes net representation of Figure 1(a).

Interpreted signals are defined by Hong and Page [2009] to produce predictions about outcomes, based on interpretations of partial observations of the underlying state. At a coarse grain, the interpreted signal scheme is represented by the Bayes net of Figure 2(a). A key distinction from the generated-signal model is that whereas the signals may be conditionally independent given $\omega$, they are not structurally separated by the outcome variable $v$. This means that if we attempt to transform the graph so that the distribution of $s_i$ is expressed in the form of a conditional on $v$ (e.g., by marginalizing the $\omega$ variable), as in generated models, we would introduce direct probabilistic dependencies among the signals.

The more detailed framework for interpreted signals developed by Hong and Page [2009] is presented in Figure 2(b). Although the original work did not employ Bayes nets, we describe here how the graphical modeling formalism naturally captures the structural assumptions defined by that scheme.

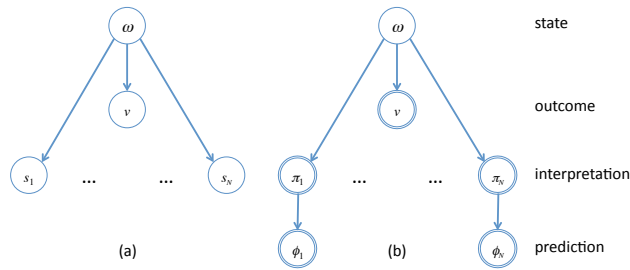

Figure 2: (a) Basic pattern of interpreted signals. (b) Structured framework for interpreted signals presented by Hong and Page [2009].

Interpreted signals in this framework require that the outcome variable depends deterministically on the underlying state. This is expressed in the Bayes net picture by the double circle employed for the node $v$. The private information or *interpretation* $\pi_i$ of agent $i$ is also a deterministic function of underlying state, which entails that the possible signals define a partition over these states. Since these relations are deterministic, we use $v(\omega)$ to denote the value of outcome variable $v$ given state $\omega$, and similarly $\pi_i(\omega)$ for the interpretation of agent $i$ given that state.

Given its interpretation, agent $i$ forms a *prediction* $\phi_i$ of the outcome variable by selecting its most likely value:

$$\phi_i(\pi_i) = \arg\max_v \sum_{v(\omega)=v \land \pi_i(\omega)=\pi_i} \Pr(\omega).$$

We can also define predictions as functions of underlying state, $\phi_i(\omega) \equiv \phi_i(\pi_i(\omega))$.

A further hallmark of the interpreted framework is its appeal to a representation of the state in terms of *attribute* or feature vectors. Interpreted signals are based on observing a subset of attributes. For example, an interpreted signal for the value of a car might comprise the age, mileage, type, and condition of the car. A particular realization of this signal (i.e., the interpretation) corresponds to an equivalence class of possible cars that share the same attribute values. The agent's prediction for car value (the outcome) is that value maximizing total probability of cars consistent with both the value and the signaled attributes. This prediction is imperfect to the extent the interpretation does not consider all relevant attributes.

An attribute-based representation renders even more compelling the expression of dependence structure using graphical models. All we need to accommodate this structure is to replace the $\omega$ node in the models depicted above with a set of nodes $\{X_1, \ldots, X_K\}$ representing the respective attributes. We then add edges among these nodes to express probabilistic dependencies among these attributes.

Thus we see that Bayes nets can express both patterns of signals—generated and interpreted—and that the graphical structure itself reveals which pattern holds sway.

**Observation 3.** *Graphical dependence structure captures the central difference between the patterns of generated and interpreted signals, as defined by Hong and Page [2009].*

In typical applications, generated signals are assumed either to be independent or independent conditional on outcome. For instance, in the generated-signal model of Figure 1(a), signals are conditionally independent given value. That is not the case for interpreted signals. This too is immediately apparent from inspection of the interpreted signal structure of Figure 2(a), or 2(b). The outcome variable does *not* d-separate (see §2.2) the signals (interpretations) in these models, and thus we should not expect conditional independence. Though the lack of d-separation does not necessarily entail actual dependence, in fact, for the case of binary signals Hong and Page [2009] show that dependence is almost inevitable. Formally, they show that there exists a unique outcome function, up to isomorphism, for conditionally independent interpreted signals, and even more limiting, that for that function, each of the signals must see each attribute but one, and the missing attribute must differ for each signal. This is a very unrealistic case.

Finally, we wish to emphasize that the point of distinguishing interpreted from generated is not to argue for one model of signals over the other. To the contrary, there are classes of natural situations best rendered in terms of generated signals, and others better modeled as interpreted. Moreover, the flexible graphical representation suggests the possibility of hybrid cases, where private information includes combinations of generated and interpreted signals.

### 3.2 Independence of Interpretations

The attribute representation of state facilitates reasoning about dependence across interpretations. The interpretations of agents $i$ and $j$ are independent, if for all values of $\pi_i$ and $\pi_j$, $\Pr(\pi_i, \pi_j) = \Pr(\pi_i)\Pr(\pi_j)$. Such independence implies that knowing one agent's interpretation provides no information about the other's: $\Pr(\pi_j \mid \pi_i) = \Pr(\pi_j)$. From the Bayes net of Figure 2(b), we see that predictions are d-separated iff their associated interpretations are as well. Given the graph structure, independence of interpretations entails independence of predictions. The converse does not hold, however, as it is possible to construct scenarios where predictions turn out to be statistically independent even when interpretations interdepend [Hong and Page, 2009].

Using the graphical model formalism, we can characterize independence of interpretations directly in terms of independence among the attributes on which these interpretations are based. Figure 3(a) depicts the situation where $N$ agents have independent interpretations, because the state is factored into $N$ independent variables representing the attributes observed by the respective agents. This captures the argument by Hong and Page [2009] that a product-space representation of state is the only way to support interpretation independence. It also follows that mutual independence of nontrivial interpretations is possible only for a state space sufficiently rich to be factored $N$ ways: $|\Omega| \geq 2^N$.

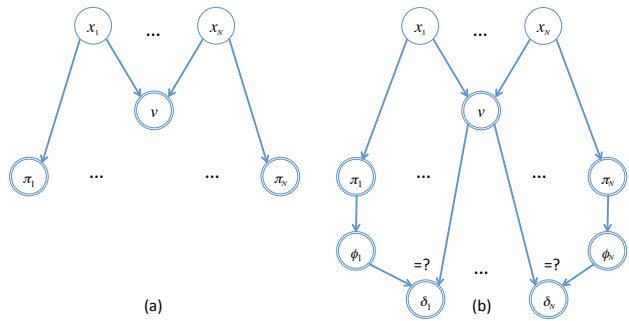

Figure 3: (a) Unconditionally independent interpretations. (b) Dependencies of prediction correctness even for independent predictions.

**Observation 4.** *The attribute-based perspective of interpreted signals is directly reflected in graphical model structure, which supports reasoning about interpretation independence.*

### 3.3 Correctness of Predictions

The prediction associated with an interpreted signal is *correct* if it coincides with the outcome that is actually the case. We can express correctness in another random variable, $\delta$, also a function of underlying state:

$$\delta(\omega) = \begin{cases} 1\ (correct) & \text{if } \phi(\omega) = v(\omega), \\ 0\ (incorrect) & \text{otherwise.} \end{cases}$$

The *accuracy* of an interpreted signal $\phi$ can then be defined as the probability of its associated $\delta$ taking value 1. We say that two predictions are independently correct if knowing that one agent's prediction about the outcome is correct gives no information about whether the other's prediction is correct. From Figure 3(b), it is clear that the correctness of predictions is dependent (by virtue of the path through $v$), even if the predictions themselves are independent. This structural dependence bears out another result of Hong and Page [2009]. For binary signals, they show that independent and informative (probability of correctness greater than 1/2) predictions that predict two outcomes with equal probability are negatively correlated in their correctness.

## 4 Signal Structure in Prediction Markets

The previous section showed that the distinction between generated and interpreted signals manifests in qualitatively different belief patterns (e.g., dependence in prediction accuracy), and thus may entail significant implications for strategic behavior. These distinctions are well captured by Bayesian networks, a property also noted in the example of interdependent value in auctions (§2.3), which also illustrates strategic implications of independence structure.

In this section we examine another scenario, about bluffing behavior in prediction markets. We use graphical models to explicate the results of Chen et al. [2010], which provide another example of strategic behavior hinging qualitatively on signal dependence structure. The authors consider the behavior of bettors in a prediction market based on a *market scoring rule* [Hanson, 2003]. The mechanism maintains a current probability distribution (suitably initialized). Participating agents may then change the distribution to one they specify, and receive a payoff based on their chosen change and the outcome of some uncertain event. The payoff is calculated using a proper scoring rule, which means that the agent with only one opportunity to bet optimizes its expected payoff by specifying its true probability distribution. The strategic question raised by Chen et al. [2010] is whether truthful reporting remains optimal when agents have the opportunity to make multiple bets over alternating rounds. The potential is opened for *bluffing*, where an agent misreports its probability initially, perhaps to mislead the others and capitalize in a later round.

The question is interesting only if the agents' signals are informative, in which case they must all be related to the outcome variable of interest. Chen et al. [2010] explore two basic signal structures, illustrated for the case of two agents in Figure 4. The first (a) is an instance of the generated signals model—signals are dependent, but conditionally independent given the outcome variable. In the second (b), signals are unconditionally independent, but dependent conditional on the outcome. Like the interpreted model, signals are viewed more as causes than effects of the variable of interest.

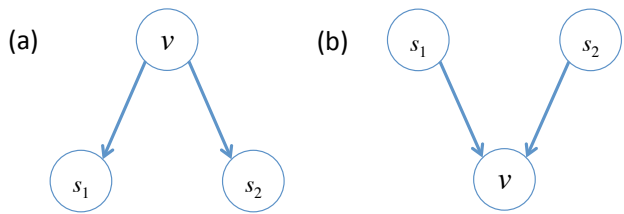

Figure 4: Signal structures for betting in prediction markets. (a) Conditionally independent signals. (b) Unconditionally independent but conditionally dependent signals.

The main result of Chen et al. [2010] is that truthful bidding is a perfect Bayesian equilibrium given the generated signals model of Figure 4(a). That is, bluffing is not profitable given that signal structure. For unconditionally independent signals, however, truthful behavior is not in equilibrium, and bluffing can be profitable. By way of intuitive explanation, the authors show that for this mechanism under model (b), the signals are *complements*: the expected profit for knowing both signals is greater than the sum of expected profits from each individually. If bets are truthful, the agent who bets second effectively gets to observe both signals. This can explain why complementarity of signals motivates bluffing. Under the generated model (a), in contrast, signals are substitutes (independent draws conditional on true outcome), and hence it is more important to get the benefit of the first signal than to hold out to exploit the combination.

## 5 Qualitative Reasoning

The idea of qualitative probabilistic reasoning [Parsons, 2001] is to capture high-level relationships among uncertain variables at an appropriate level of abstraction, relying on precise assessments of probability values only when strictly necessary for the purpose at hand. Qualitative relationships typically correspond to basic causal information, such as "$a$ positively influences $b$". These relationships may also be expressed conditionally, and capture higher-order relationships, such as "$a$ and $b$ are synergistic in their influence on $c$".[3] Such relationships tend to be quite robust, insensitive to the precise measurement scale of the variables, and often to which details are included or not included in a model. They are easier to elicit than precise numeric relationships, because they focus on causal structure and relative judgments, whereas assessing absolute qualitative measures generally requires an exhaustive consideration of contributing factors. Moreover, qualitative models admit efficient inference algorithms based on sign propagation, in contrast to the worst-case exponential calculations required for exact numeric queries.

The qualitative probabilistic network (QPN) formalism [Wellman, 1990] captures basic influence and synergy relationships, described informally above,[4] and were shown to capture important patterns of reasoning, such as tradeoff formulation, "explaining away" [Wellman and Henrion, 1993], and other intercausal patterns [Lucas, 2005]. Druzdzel and Henrion [1993] developed computationally efficient inference algorithms based on qualitative belief propagation. Vanderweele and Robins [2009] show how qualitative relations can be used for causal assessment of statistical tests. Additional research has extended QPNs in various ways, to improve expressive power [Renooij et al., 2002] and provide some basis for tradeoff resolution [Liu and Wellman, 1998, Renooij and van der Gaag, 1999].

---

[3]Conditional independence may itself be viewed as a kind of qualitative relation, and in that sense the annotations discussed here are a natural extension to graphical model representations.

[4]See cited references for precise technical definitions in terms of probabilistic inequalities.

Qualitative probabilistic relations capture essential distinctions in the economic analysis of signals. Indeed, the concept of positive probabilistic influence in QPNs is based on the same monotone likelihood-ratio property employed by Milgrom [1981] to categorize signals as "good news" or "bad news". Similarly, the synergy relations defined by QPNs are related to qualitative properties used by economists in various domains (auctions, oligopoly, signaling games) to establish that specified game classes have equilibria in (pure) monotone threshold strategies [Athey, 2001, Milgrom and Shannon, 1994].

We illustrate the use of qualitative reasoning for games of incomplete information with a simple auction example. Figure 5 presents a QPN[5] for a single-item first-price sealed-bid (FPSB) auction with two bidders. The left side of the diagram is exactly the generated signal model of interdependent values (Figure 1(a)), augmented with "+" signs indicating the direction of probabilistic relations. Intuitively, the edge $\omega \to^+ v_i$ says that value is positively related to the underlying state (this requires that the state variable have some ordered scale), and $v_i \to^+ s_i$ that the signal $s_i$ is likely to be higher for higher $v_i$ values. The positive path between the two signals (influences combine by sign multiplication) means they are *affiliated* [Milgrom and Weber, 1982], which is the fundamental assumption guiding many results in the theory of auctions with interdependent value [Krishna, 2010, Milgrom, 2003].

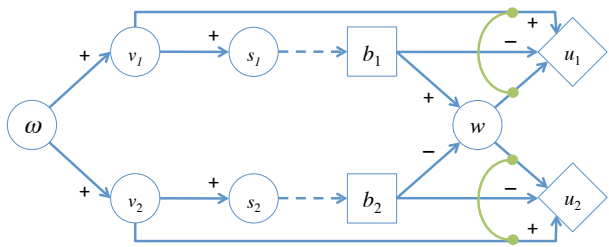

Figure 5: Qualitative probabilistic model of a two-bidder FPSB auction with interdependent values.

The remaining variables represent:

- the bid $b_i$, drawn in squares to indicate decision nodes,
- the "winner" $w$, based on comparing the bids, and
- the utility values $u_i$, drawn in diamonds to indicate value nodes.

The dashed arrow from $s_i$ to $b_i$ indicates that agent $i$ knows the value of $s_i$ (i.e., has observed the signal) at the time it decides its bid. For current purposes we take the perspective of agent 1, and interpret the variable $w$ as *true* iff agent 1 wins (i.e., $b_1 > b_2$, or with probability 0.5 if $b_1 = b_2$.) Under the convention $true > false$, there is a positive probabilistic influence from $b_1$ to $w$, and a negative influence from $b_2$ to $w$. The utility for agent $i$ is $v_i - b_i$ if it wins, zero otherwise, hence $v_i$ is a positive influence on $u_i$ and $b_i$ negative. The influence of $w$ is ambiguous, given the uncertainty about whether $v_i \geq b_i$.[6]

The last key element of the qualitative model is indicated by an arc connecting the influences of $v_i$ and $w$ on $u_i$. This notation represents the positive *qualitative synergy* (a probabilistic form of superadditivity [Wellman, 1990]) among these variables. Intuitively, value is synergistic with winning on utility because the effect of an increase in value is greater when the agent wins (in this case trivially, as value has no influence if the agent loses).

From the qualitative relations specified, we can infer an important (and well known) fact about agent strategy: that an agent's optimal bid is monotone increasing in its signal. This follows from the monotone policy implications of positive synergy between the bid and signal. We infer this in turn from (1) the positive path from signal to value, (2) the positive path from bid to win, and (3) given synergy of value and win. This inference is not affected by the interdependent values, and would go through as well in the IPV model with the state variable $\omega$ omitted.

What does rely on interdependent value is the inference we can make about the influence of winning on value.

**Definition 2** (Winner's Curse). *An auction model exhibits winner's curse if the event of winning has a negative qualitative probabilistic influence on the agent's value.*

We incorporate the monotone policy conclusion above by representing $b_2$ (from agent 1's perspective) as a random variable with positive influence from $s_2$. With this modification of the model, we find a negative path from $w$ through $b_2$ to $v_1$. This means that for any chosen bid $b_1$, the event of winning is bad news about the good's value. Thus the infamous winner's curse is here explicated through a graphical model with qualitative annotations.

Since bid $b_1$ is synergistic with value $v_1$ (same inference as for the signal, but more direct), the winner's curse is a reason to bid lower in FSPB auctions. Eliminating the underlying state $\omega$ in this case would break the inference path; winning is no curse in the IPV model. It follows that the interdependence of value lowers bids in FPSB, all else equal.

**Theorem 5.** *The QPN model of FPSB (Figure 5) entails*

---

[5]Technically since there are multiple agents we should call this a qualitative *MAID* [Koller and Milch, 2003].

[6]The extension of QPNs to allow context-specific signs [Renooij et al., 2002] would support more precise inference in this case, since the influence of $w$ is disambiguated by the other predecessors of $u_i$.

*monotone bidding strategy, winner's curse, and the reduction of bids to compensate for winner's curse.*

*Proof:* The results follow from standard QPN definitions and inference rules [Druzdzel and Henrion, 1993, Wellman, 1990], as sketched above. □

Note that the reasoning above goes through as long as the signals are affiliated (have a net positive path in the qualitative model), and also positively related to value. The conclusions about winner's curse do not depend on this specific generated-signals model. We can construct instances of interpreted-signal patterns, for example, where the signals are not mutually affiliated with value, but are mutually affiliated with each other and positively related to value.[7]

**Observation 6.** *Winner's curse properties (and other results of interest) can be established from interpreted signal patterns, not covered by the standard theorems.*

Through this sort of analysis it appears possible to generalize many existing auction-theoretic characterizations from the literature, using an expanded repertoire of qualitative and structural conditions.

A QPN model of the second-price sealed-bid (SPSB) auction would be similar to Figure 5: we need only replace the $b_i \to u_i$ edges with edges pointing to the other agent's utility node, $b_i \to u_{-i}$. The inferences relating signals to bids and winner's curse stay valid, though the implications of winner's curse are quite different. The event of winning is still bad news for *value* under SPSB, but a low other-agent bid is also good news for *utility*, as it lowers the price the winner pays. Which influence holds sway depends on the particulars. We have constructed an example [Wellman, forthcoming] where a bidder with a high signal in SPSB auctions bids *higher* given interdependent values than it would with independent values, and a bidder with a low signal bids lower.

## 6 Conclusion

Graphical models provide a language for describing and reasoning about the qualitative structure of a decision or game situation, including the private signals on which agents base their decisions. By explicating the pattern of signals, we gain insights about abstract properties of these games. In this paper we have identified some such properties reflected directly in graphical structure. In particular, we have shown how the framework of interpreted signals developed by Hong and Page [2009] can be recast in terms of distinctions on graphical structure, and argue that this representation facilitates reasoning about its implications. The perspective also captures pivotal distinctions in strategic analysis of prediction markets, and auctions. Qualitative models of auction games that augment independence structure with monotonicity relationships capture important patterns of reasoning in these domains, and suggest generalizations on known results.

Graphical models may also be used more concretely, for solving instances of incomplete-information games specified using this formalism. If we augment the directed graph with specifications of precise conditional probabilities at each node, we have a complete Bayesian network representation, and can employ techniques from a rich algorithmic toolkit for answering queries about probabilistic relationships in the model [Koller and Friedman, 2009]. Extensions to include decisions and utilities of multiple agents yield a graphical model of a game situation. Koller and Milch [2003] call this extended model form a *multiagent influence diagram* (MAID) and show how to utilize both graphical model and game-solving computational techniques for inference in the MAID representation. Gal and Pfeffer [2008] extend the possibilities even further, with their *network-of-influence-diagram* (NID) representation. NIDs provide a straightforward way to express that agents may have differing views of the game, by defining each agent's view of the game as its own MAID.

In future work, we plan to develop canonical Bayes-net model structures for signal patterns, expressed as building blocks within MAIDs or NIDs, thus facilitating more widespread use of these models for analyzing incomplete-information games. Another promising direction is to explore models that *endogenize* signals, by explicitly representing the decision to observe outcome-relevant attributes. We anticipate that understanding qualitative relations among signal attributes will be essential for characterizing the form of optimal observation policies.

## A  Example: Affiliation with Interpreted Signals

The seminal paper by Milgrom and Weber [1982] established several auction-theoretic results under the assumption of *affiliation*: essentially, that the higher one agent's signals, the more likely that other agents' signals and values are to be higher as well.

We demonstrate by example that the affiliation requirements of interpreted signals may be more flexible than the standard generated signal model. Let there be two bidders and a common value object. The common value, denoted by $v$, can take two possible values, 0 or 1. Prior to bidding, each agent gets a signal $s_i \in \{0, 1\}$. The joint distribution of signals and the value is described as follows. Each of the configurations $(s_1, s_2; v)$: $(0, 0; 0), (1, 0; 1), (0, 1; 1)(1, 1; 1)$ has probability 0.25, and the remaining combinations have zero probability. Notice that $\Pr(0,0;1) \cdot \Pr(1,1;1) = 0 < \frac{1}{16} = \Pr(1,0;1) \cdot \Pr(0,1;1)$. This construction therefore violates

---
[7]See Appendix A.

affiliation. (Another way to see this is that the signals are negatively correlated given $v = 1$.)

If we treat the signals as interpreted, however, then it becomes clear that the signals themselves are unconditionally affiliated. To show this we first write the relationship between signals and the common value given by the joint probability distribution of signals using an outcome function that depends on two attributes $x_1$ and $x_2$:

$$v = f(x_1, x_2) = x_1 + x_2 - x_1 x_2,$$

where $x_i$ can either be 0 or 1 with equal probability and $x_1$ and $x_2$ are independent. Prior to bidding, bidder $i$ observes the signal $s_i = x_i$ and bids accordingly.

Since $s_1$ and $s_2$ are independent, they are trivially affiliated. A qualitative model incorporating this situation for a common-value FPSB is depicted in Figure 6. The conclusions about FPSB strategy stated as Theorem 5 likewise follow for this configuration, by the same series of inferences applied there. Affiliation of signals in conjunction with the positive relation between signals and value is also enough to establish the results about SPSB equilibrium reported by Milgrom and Weber [1982].

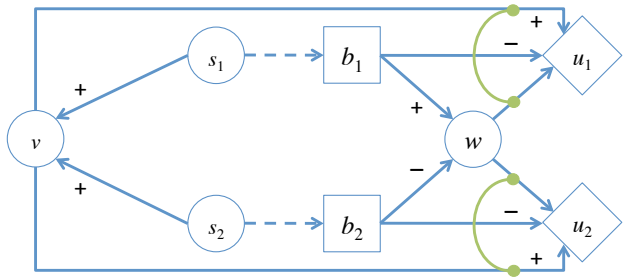

Figure 6: Qualitative probabilistic model of a two-bidder FPSB auction with common values and interpreted signals.